# Fast Membranes Hemifusion via Dewetting between Lipid Bilayers

Jose Nabor Vargas [a], Ralf Seemann [a,b] and Jean-Baptiste Fleury [a,*]

**ABSTRACT**

The behavior of lipid bilayer is important to understand the functionality of cells like the trafficking of ions between cells. Standard procedures to explore the properties of lipid bilayer and hemifused states typically use either supported membranes or vesicles. Both techniques have several shortcoming in terms of bio relevance or accessibility for measurements. In this article the formation of individual free standing hemifused states between model cell membranes is studied using an optimized microfluidic scheme which allows for simultaneous optical and electrophysiological measurements. In a first step, two model membranes are formed at a desired location within a microfluidic device using a variation of the droplet interface bilayer (DiB) technique. In a second step, the two model membranes are brought into contact forming a single hemifused state. For all tested lipids, the hemifused state between free standing membranes form within hundreds of milliseconds, i.e. several orders of magnitude faster than reported in literature. The formation of a hemifused state is observed as a two stage process, whereas the second stage can be explained as a dewetting process in no-slip boundary condition. The formed hemifusion states are long living and a single fusion event can be observed when triggered by an applied electric field as demonstrated for monoolein.

## Introduction

Cell membrane fusion is an important cellular process that occurs when two initially separated lipid membranes merge into a single continuous bilayer. Membrane fusion plays an important role in a wide range of physiological processes as cell growth, membrane repair, cytokinesis, intracellular transport or synaptic transmission[1]. The fusion of membranes is typically driven by proteins (SNAREs, Hemagglutinin)[1-6] or by divalent cations in the case of synthetic vesicles[7-9]. But despite of a large number of studies during the last 30 years, many aspects of membranes fusion stayed controversial like the pathways to fusion in the case of protein-free lipid bilayers[1-10]. Several studies are suggesting that the pathway to fusion begins with an initial hemifused region of minimum size (stalk), where the outer leaflets of the opposing membranes are fused while the inner leaflets engage in a new bilayer region called the hemifusion diaphragm (HD)[1-11]. Subsequently, the HD region is expanding until an equilibrium hemifused state is reached, unless the (HD) membrane tension is sufficient to trigger the membrane fusion[7-9]. The actual lack of knowledge which might be partly due to the lack of suitable experimental techniques to investigate single fusion event[1-10].

In this work, we present a microfluidic platform which can be used to produce and to study the different stages of membrane fusion between two free standing lipid bilayers. The microfluidic platform is based on the concept of droplet interface bilayer (DiB)[12] which was proposed by H. Bayley et al. and allows in a first step to generate stable free standing lipid bilayers. In a second step, the two generated bilayers can be manipulated and e.g. brought into contact to each other. Using this microfluidic scheme we characterize the formation of free standing hemifused states from previously formed free standing bilayers by simultaneous optical and electrophysiological measurements. The observed formation of a hemifused state is surprisingly fast and can be characterized as a two stage process, whereas the second stage can be identified as dewetting process between molecular thin layers. We also study a single fusion event of an initially stable free standing hemifusion diaphragm by electrofusion. The found experimental results for free standing bilayer and membranes seem to deviate from results obtained by other methods and support the need for alternative experimental approaches.

## Results and Discussions

We will first describe the formation of free standing membranes and hemifusion states. Subsequently, the formation of membranes of various lipids and in particular of hemifused states formed from these membranes is discussed. The formation process of hemifused states is identified as a two stage process, whereas the second stage is explained as a dewetting process. The electrofusion of a single hemifusion state is analyzed and discussed for monoolein.

## Quasi-Automated Formation of a Single Equilibrium Hemifused State

The used microfluidic devices are made of Sylgard 184 (Dow Corning) and consists of two hydrophobic microchannels in a cross geometry, see figure 1. The liquid flows are volume controlled using syringe pumps and the entire process is monitored by optical microscopy and potentially supplemented by simultaneous electrophysiological measurements. In a first step, two water fingers are injected face-to-face into the cross geometry which was previously filled with an oil-lipid solution (figure 1.a). After a few seconds, the water-oil interface of each finger is covered with a monolayer of lipid molecules[12-13]. When two such liquid fingers are brought in contact, the two lipid monolayers are interacting and form a lipid bilayer[12-13] within a short time. Similarly, two bilayers are formed when slowly injecting a third water finger into the middle channel, as shown in figures 1.b,c. The third finger is replacing most of the oily phase which was initially present in the middle channel and two parallel lipid bilayers are formed at each liquid interface (supporting movie 1). Then the two lipid bilayers can be approached by injecting more liquid into the initially injected aqueous fingers. When the two bilayers get in contact, they automatically trigger the formation a long-lived hemifusion state after a few seconds (figure 2.a).

# Characterization of the Fast Formation of a Single Equilibrium Hemifused State

**Characterization of the Fast Formation of a Single Lipid Bilayer**

When two lipid covered oil-water interfaces are brought in contact they form a bilayer, as described in the previous section. However, it is important to note that this bilayer formation is not immediate but emerges from a two-step process[13]. Figure 3 shows the capacitance measurements recorded during the formation of a monoolein membrane: When the two monolayers are first brought into contact, a contact area is formed having a capacitance value of C ~ 17 pF. After a few seconds in contact, the capacitance value suddenly increases (within about 250 ms) to C ≈ 200 pF. Subsequently to the sudden increase a further slow increase of the capacitance signal is observed which corresponds to a further increase in contact area. The area increase comes to a hold when the geometric limitations of the microfluidic device are reached.

Using the optically measured contact area, the specific capacitance at initial contact can be calculated to $C_s \approx 0{,}54$ mF/m². With the dielectric constants of the lipid membrane and the oily phase (squalene/decane) $\varepsilon_L \sim \varepsilon_{oil} \sim 2{,}2$[12-13] and the vacuum permittivity, $\varepsilon_o \approx 8.85 \cdot 10^{-12}$ F/m, the thickness of the contact area $d \approx \varepsilon_0 \varepsilon_L / C_s \approx 36\ nm$ can be calculated from the specific capacitance. The thus obtained initial thickness of the contact area is about one order of magnitude larger than the typical thickness of a monoolein membrane (~3 nm)[12-13] suggesting that an oil film of about 33 nm is still present between the two monolayers at initial contact.

The corresponding thickness of the contact area after the sudden increase of the capacitance value can also be calculated from the corresponding specific capacitance, $C_s \approx 6{,}3$ mF/m², and yields about 3 nm for the used monoolein lipids. This value is in agreement with monoolein bilayer thickness reported in literature[13-14,19], meaning that the two monolayers have self-assembled into an oil-free bilayer during the sudden capacitance increase. The fast drainage of the initially present oil nanofilm can be observed simultaneously with optical high speed microscopy, see figure 4 (supporting movie 3). The displacement of the oil always starts from one side of the contact area and moves like a zipper with constant velocity towards the other side. A start of the monolayer-zipping somewhere in the middle of the contact area was never observed. The velocity of this membrane zipping can be extracted from the optical measurements and is plotted for four different types of lipids (DOPC, DPhPC, monoolein and POPC) in the bottom tile of figure 4. It is interesting to note that the observed *zipping* velocities of DOPC, POPC and monoolein scale with an accuracy of about 20 % with the membrane tension, $\Gamma = 2\gamma \cos \theta$, where $\gamma$ is the surface tension of the oil/water interface and $\theta$ is the contact angle of a single membrane[14]. In contrast, the zipping velocity observed for a DPhPC bilayer clearly escapes from this scaling and it seems that the zipping velocity does not scale generally with the membrane tension, $\Gamma$. The comparison of the various determined values can be found in Table 1.

The above described membrane formation process is highly reproducible and a solvent free bilayer can be safely assumed when the initial two monolayers are in contact for times longer than a few seconds. The detailed analysis of the membrane formation mechanism, however, is beyond the scope of this article.

**Fast Formation of a Single Hemifused State**

Following the protocol described in the previous section, two parallel bilayers are produced. In a second step, the two bilayers are brought into contact moving the outer liquid filaments towards the center of the device. The entire process is observed by direct optical visualization and by simultaneous capacitance measurements. To guarantee a measured capacitance signal from only the contact area and to avoid any undesired signal from the other membranes being separated by a water layer, the water finger injected from the side channel contains ultrapure water free from any salt, see inset of figure 5. The capacitance signal obtained during the first seconds after converging two DPhPC membranes is displayed in figure 5. At t ~ 0 s, the two bilayers are touching each other as determined from optical measurement and the total capacitance is equal to C ≈ 40 pF. This capacitance value stays stable until t ~ 6.5 s, when a spontaneous increase of the capacitance value to C ~ 200 pF is observed. The capacitance value further increases slowly due to an increase of contact area reaching a final value.

Using the optically measured contact area, the specific capacitance at initial contact can be calculated to $C_s \approx 0{,}75\ mF/m^2$. The layered system consists of a thin water layer sandwiched between two solvent free bilayers. To calculate the thickness of the layer, an equivalent circuit of three dielectrics in series is considered. Using the dielectric constants of the involved lipids $\varepsilon_L = 2{,}2$ and of pure water $\varepsilon_{water} = 88$, yields $d \approx \left(1/C_s - \frac{2d_{DPhPC}}{\varepsilon_0 \varepsilon_L}\right) \varepsilon_0 \varepsilon_{water} \approx$ 0,64 µm, indicating a water microfilm between the two oil-free bilayers ($d_{DPhPC} \sim 5\ nm$)[12]. The specific capacitance after the sudden increase is constant and can be calculated to $C_s \approx 3{,}78$ mF/m² revealing a reduced thickness of only ~ 5 nm of the contacted DPhPC membranes which corresponds to the thickness of a pure DPhPC membrane[19]. This demonstrates that the analyzed structure which emerges at *t* ~ 6.5 s after contacting the two DPhPC membranes is a hemifused state[11,15-19,22]. This hemifused state is long-living, and can be stabilized for more than 1 hour. The lifetime is limited only by coalescence of the aqueous fingers due to drainage of the continuous oil phase into the Sylgard 184[20-21] and could be extended by replacing the Sylgard 184 with a non-porous material like glass. Moreover, no rupture of the contact area of the membranes (i.e. fusion event) could be observed without external stimulation. From these findings we conclude that this state of hemifusion can be considered as an equilibrium state[15,16].

**Dynamics of HD Expansion by Simultaneous Optical and Capacitance Measurements**

In the following we explore the above explained, fast formation of a hemifusion diaphragm in more detail using simultaneously recorded optical high speed micrographs. When the capacitance signal increases and the hemifusion starts ($t \sim 6.5$ s in fig. 5), a symmetry breaking is observed optically at a random position within the contact area of the two bilayers (supporting movie 4a). From the observation angle parallel to the contact area, the symmetry breaking resembles a localized travelling wave which is reflected several times within an area of 20 - 60 µm diameter. During these fluctuations, an initial HD region with a diameter of 20 – 60 µm is formed within ~ 10 ms and the capacitance signal increases from ~ 40 pF to ~ 100 pF. Not that the formation of initial HD area can only be seen in the optical signal and is not resolved by a particular feature in the capacitance measurement. However, assuming the just formed initial HD area to have the thickness of a bilayer, ~ 5 nm, the thickness of the remaining water layer can be calculated to about ~ 150 nm. It is remarkable that the initially thickness of the water layer is reduced by about 500 nm during this short period of ~ 10 ms. This drainage process is about two orders of magnitude faster than the drainage of oil between two monolayer (table 1 and figure 4). This drainage is also significantly faster that the further growth in HD area, as will be explained next. The different dynamics may indicate a different physical drainage mechanism between the zipping and the hemifusion processes.

We define $\Delta t \sim$ 0 ms (in figure 6 and 7) when the initial HD region is formed. At this time distinct structures emerge at the boundaries of the just formed initial HD area (contact lines) which travel both outwards from this area, see figure 6. The increase in radius of the travelling contact lines is very fast ~1 mm/s and linear in time ($\Delta R \sim \Delta t^{1,1}$) for all tested lipids. When reaching the geometrical limits of the microfluidic device after another $\Delta t \sim$ 25 ms, the contact line velocity is reduced and finally comes to a halt. The observed dynamics of the drainage of the water layer indicates that the expansion of the HD region is driven by a dewetting process with no-slip boundary conditions[23-24]. This would mean that in a first stage a HD area is formed and the emerging contact lines are driven by a minimization of the involved interfacial energies. The corresponding dewetting velocity, $\frac{d\Delta R}{dt}$, should thus scale linearly with the spreading coefficient, i.e the membrane tension $\Gamma_{HD}(\cos\theta_{HD} - 1)$. For an hemifused membrane, the membrane tension $\Gamma_{HD}$ can be expressed as $\Gamma_{HD} = 2\,\Gamma \cos\theta_{HD}$. Here, $\Gamma$ is the membrane tension of a single bilayer which can be calculated from the measured surface tensions of lipid decorated oil/water-interfaces and $\theta_{HD}$ is the contact angle of the hemifused membrane which can be determined from optical micrographs. Using the thus determined values confirm that the drainage velocities, $\frac{d\Delta R}{dt}$, for all studied lipids scale in fact linearly with the spreading coefficient within a accuracy of about 15%, i.e. in quantitative agreement with experimental accuracy.

Based on this model, i.e. expansion of the HD area by a dewetting process, see figure 7.a, the increase in hemifused area $A_{HD}$ during the dewetting process can be calculated from the capacitance measurements. The equivalent circuit consist of the non-dewetted area and the dewetted area in parallel, see figure 7.b. The non-dewetted area can be represented by three dielectrics in series (lipid bilayer, water, lipid bilayer) and the dewetted area is represented by one dielectric (lipid bilayer):

$$A_{HD}(t) = \frac{A_{total}B - \frac{C(t)}{\varepsilon_0 \varepsilon_L}}{B - \frac{1}{d_{DPhPC}}},$$

with $\quad B = \frac{\varepsilon_{water}}{2d_{DPhPC}\varepsilon_{water} + d_{water}\varepsilon_L} \quad$ (1)

For a dewetting process, the dewetted water layer accumulates in a rim which is localized at the contact line and the remaining water thickness can be assumed as constant in thickness. The extension of the rim can be assumed as small compared to the extension of both the dewetted and the non-dewetted area and is disregarded in the calculation. Assuming the thickness of the non-dewetted area ($A_{total} - A_{HD}$) as constant ~150 nm (water plus two bilayers) and the thickness of the dewetted area $A_{HD}$ as ~ 5 nm (bilayer), the increase of the hemifused area $\Delta A_{HD}$ can be determined to $\Delta A_{HD} \propto \Delta t^2$ in quantitative agreement to the radial expansion expected for a dewetting process, see figure 7.c. Thus one can conclude that the observed growth of the hemifusion diaphragm after the formation of an initial HD region is driven by a dewetting process.

It is remarkable that the growth rates observed for free standing hemifusion diaphragms (~ 1 mm/s) are about five orders of magnitude larger than the growth rates observed for supported bilayers (~ 0.01 µm/s)[11]. However, considering the observed initial fluctuations and the following dewetting process it is immediately clear that the presence of a solid support will significantly influence the drainage process. Moreover, supported bilayers experience Van der Waals forces from the substrate, which additionally affect the forces acting on the membrane. The massive influence of a solid support can be also seen by comparing the results from similar measurements determining the formation of hemifused states between giant unilamellar vesicles (GUV) in bulk[22]. The observed formation speeds

(~ 4 μm/s) are just about two orders of magnitude smaller than observed here, but the increase of the HD area is linear with time. We suspect that the slower dynamics found for GUV's originates from the geometrical confinement provided by the GUV's.

## Realization and Characterization of a Membrane Fusion Event

As already mentioned, the formed hemifusion diaphragms are very stable for all tested lipids and can be considered as equilibrium or dead end states. To drive such a stable hemifusion diaphragm to fusion, certain proteins could be incorporated into the bilayers. Alternatively, a sufficiently large electric field can be applied across the hemifusion diaphragm (electrofusion), as commonly used to fuse mammalian cells or synthetic vesicles [25].

Here, electrofusion is applied as the easiest possibility to trigger the fusion of a single and stable hemifusion diaphragm. As no ionic charges are dispensed in the initially injected middle water channel, cf. figure 5, only the HD is targeted when applying an AC-voltage (figure 2.b and supporting movie 5). The measured current signal during a voltage sweep with continuously increasing voltage amplitude is shown in figure 8.a for a monoolein-HD. The signal is reversible for applied voltages up to about $U \sim 300$ mV, where only subcritical hydrophilic nanopores are appearing as a result of the applied voltage. These subcritical nanopores are too small to grow spontaneously and to fuse the membrane. The measured transported ionic current is increasing massively between $U \sim 210$ and $U \sim 300$ mV, c.f. figure 8. For applied voltages larger than about $U_c \sim 300$ mV a strongly increasing ion current is measured and the fusion of the HD is observed within milliseconds.

From the voltage dependent membrane capacitance in the reversible electroporation region, the effective membrane thickness $d$ can be determined as a function of the applied transmembrane voltage[25-27], see figure 8.b. From the effective thickness $d$, the corresponding critical radius of nanopores $r_c$ can be determined from a model based on the membrane free energy $W$[27]:

$$\Delta W(r, U) = 2\sigma \pi r - \Gamma_{HD} \pi r^2 - \frac{\varepsilon_o(\varepsilon_{water} - \varepsilon_L)\pi r^2}{2d} U^2, \quad (2)$$

where $U$ is the applied transmembrane voltage and $\sigma$ is the pore edge tension ($10^{-11}$ J/m)[25-28] and $\Gamma_{HD}$ the membrane tension ($3, 7\ mN/m$). The transmembrane voltage reduces the critical pore radius $r_c$, which is obtained by $d(\Delta W(r, U))/dr = 0$.

$$r_c(U) = \frac{\sigma}{\Gamma_{HD} + \frac{\varepsilon_o(\varepsilon_{water} - \varepsilon_L)}{2d} U^2}, \quad (3)$$

Using this approach, the critical radius for the studied monoolein-HD and for a given applied transmembrane voltage can be calculated to $r_c \sim 2.8$ nm for $U \sim 10$ mV and $r_c \sim 0.7$ nm for $U \sim 210$ mV. This last point means that the thermal energy needed to trigger the fusion at $U \sim 210$ mV is around $10 k_B T$. The minute critical pore radius at $U \sim 210$ mV explains that fusion is triggered when a voltage larger than 210 mV is applied for sufficiently long times.

Whereas the emergence of reversible nanopores in bilayer is well known, the emergence of reversible and non-destructive nanopores in a hemifused state using electroporation is remarkable and was only speculated on, e.g. Chernomordik et al[17]. However, comparing the obtained I-V characteristics of a monoolein-HD with the I-V characteristics of a monolein bilayer reported for a similar geometry[14] we find characteristic differences in their stability. The current signal for a certain applied voltage and thus the Ohmic resistance in the reversible regime is about twice as large for a monoolein-HD as compared to a monoolein bilayer. Also the HD-breakdown voltage for the monoolein HD ($U_c \sim 300$ mV) is lower than the breakdown voltage of $U_c \sim 500$ mV reported for a monoolein bilayer. This difference in stability against an applied transmembrane voltage is probably a result of the increased membrane tension in the state of hemifusion and the thus due to the reduced critical nanopore radius.

# Material and methods

**Lipid Molecules and Solutions**: Four types of lipids were used in this study, a synthetic lipid: 1,2-Diphytanoyl-sn-glycero-3-phosphocholine (DPhPC) (Avanti Polar Lipids) and three natural lipids: 1-oleoyl-rac-glycerol (monoolein) (Sigma-Aldrich), 2-Oleoyl-1-palmitoyl-sn-glycero-3-phosphocholine (POPC) (Sigma-Aldrich) and 1,2-dioleoyl-sn-glycero-3-phosphocholine (DOPC) (Avanti Polar Lipids). We have chosen these four phospholipids as these molecules are the most commonly used ones to produce and study DiB[12]. This popularity is due to their biorelevance (in particular for DOPC and POPC) and stability (for DPhPC and monoolein). To prepare the lipid solutions, 9 mg of Monoolein were dissolved in 1 ml Squalene /Decane (75 % / 25%), (Sigma-Aldrich) or 15 mg of DPhPC (or DOPC or POPC) (Avanti Polar Lipids) in 1 ml Squalene/Decane (75 % / 25%), respectively. No swelling of the Sylgard 184 (PDMS rubber) was observed using this mixture of Squalene/Decane. A 150 mM solution of NaCl (Sigma-Aldrich) in Milli-Q water was used as an electrolyte for electrophysiological measurements. The distance between the electrodes an the membrane is about a few millimeters, similarly than ref[12-14].

**Microchip Fabrication**: Microchannels with rectangular cross section were fabricated using typical soft lithography protocols. Channel dimensions were 300 μm in width and 140 μm in height. The device was molded with a SU-8 photoresist on a silicon wafer using Sylgard 184 (Dow Corning, USA). The surface of the Sylgard 184 devices was exposed to oxygen plasma (Diener electronic GmbH, Germany) and sealed with a plasma treated glass cover slide. The sealed device was rendered hydrophobic by heating it to 135 °C over night.

**Microfluidics**: The liquids were dispensed from syringes (Hamilton Bonaduz AG, Switzerland), which were connected to the microfluidic device by Teflon tubing. A homemade computer controlled syringe pump was used to control the injection of the water and oil phases, respectively. Microfluidic valves have been used to stop the flow of liquids abruptly after contacting the two monolayers, or bilayers.

**Patch Clamping**: Ag/AgCl electrodes were prepared by inserting a silver chloride wire in a borosilicate glass pipette (outer diameter 1.5 mm, inner diameter 0.86 mm, Vendor) containing an electrolyte agarose solution. Lipid membrane capacitance was measured using the Lock-In function provided by the patch clamp amplifier EPC 10 USB (Heka-Electronics). A 10 mV sinusoidal wave with a frequency of 1 kHz was used as an excitation signal. The electrodes are carefully introduced into the aqueous compartment of the Sylard 184 device. The flow and the formation of bilayer and hemifusion diaphragm were monitored using an inverted microscope (Zeiss, Axiovert 25) and a high speed camera Photron SA3.

**Surface Tension and contact angle measurements**: Surface tensions **γ** of the various lipids monolayer at oil/water interfaces were measured with the standard pendant drop method using a commercial measurement device (OCA 20, data physics) [10] (table 1) whereas the contact angle **θ** was determined from optical micrographs [10].

FIGURES

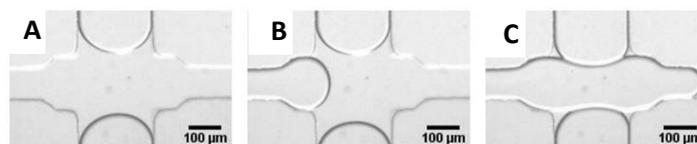

**Figure 1.** Time series showing the formation of two parallel bilayers in a cross channel geometry.

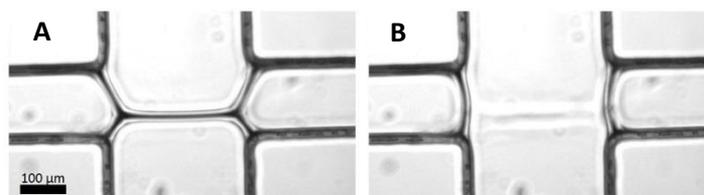

**Figure 2.** a) Hemifused membrane formed in the center of the cross geometry after approaching two initially formed bilayer. b) Remaining bilayer after electrofusion of the hemifused state shown in a).

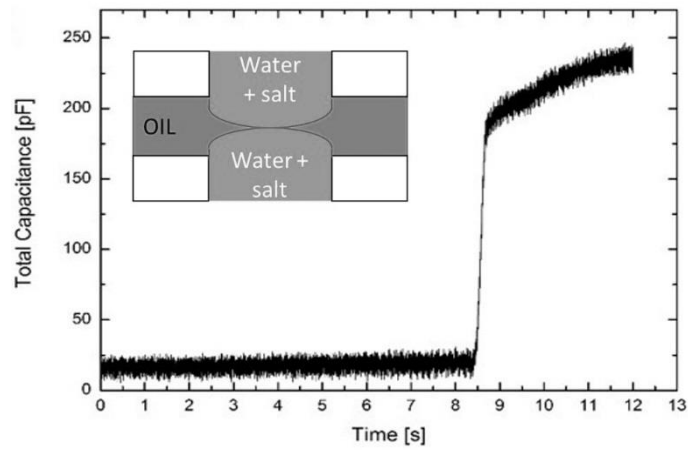

**Figure 3.** Capacitance measurement of a single monoolein membrane. *Inset:* schematic of the corresponding microfluidic situation.

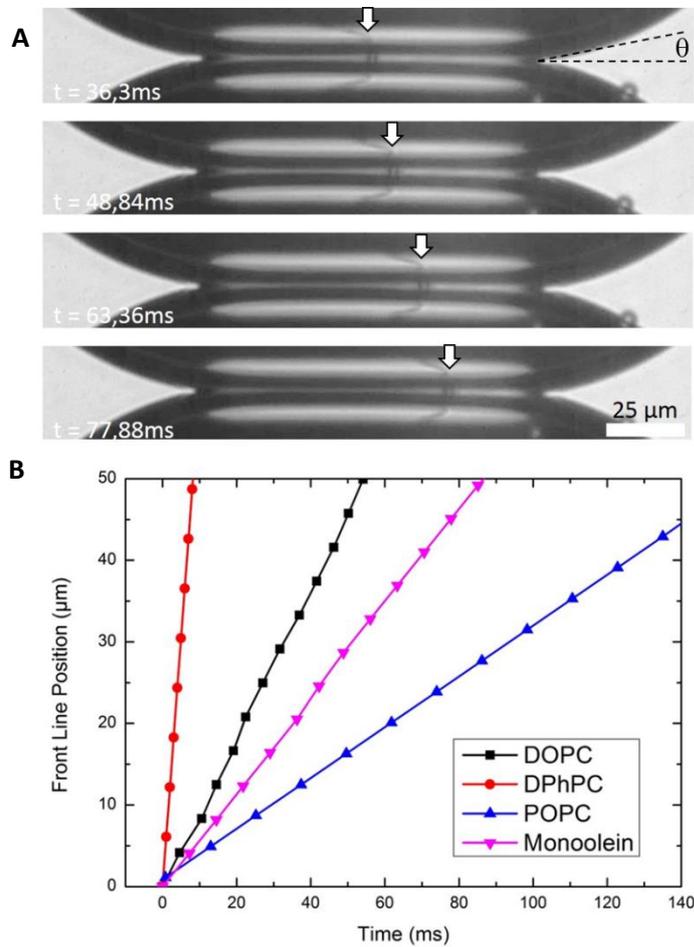

**Figure 4.** Membrane formation in the microfluidic cross geometry. *Top:* Optical time series showing a front line moving across the oil lamellae during the formation of a monoolein membrane. The white arrow indicates the moving contact line. *Bottom:* Front line position as function of time for four different types of lipids as extracted from optical time series as shown in a).

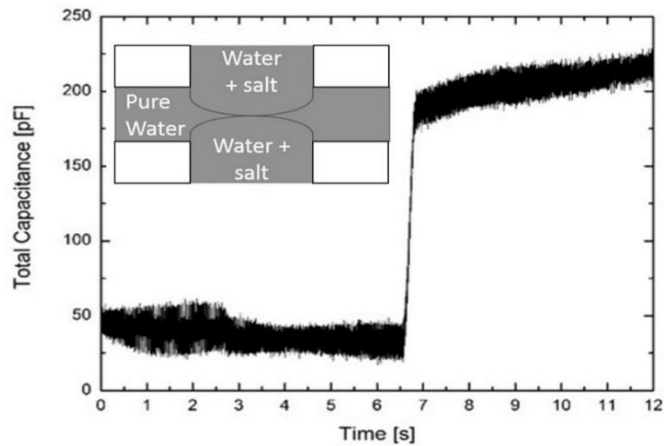

**Figure 5.** Capacitance measurement during the interaction of two DPhPC bilayers. *Inset:* schematic of the corresponding microfluidic situation: no ionic charges are present in the middle channel, the largest voltage drop is thus across the contact area in the center.

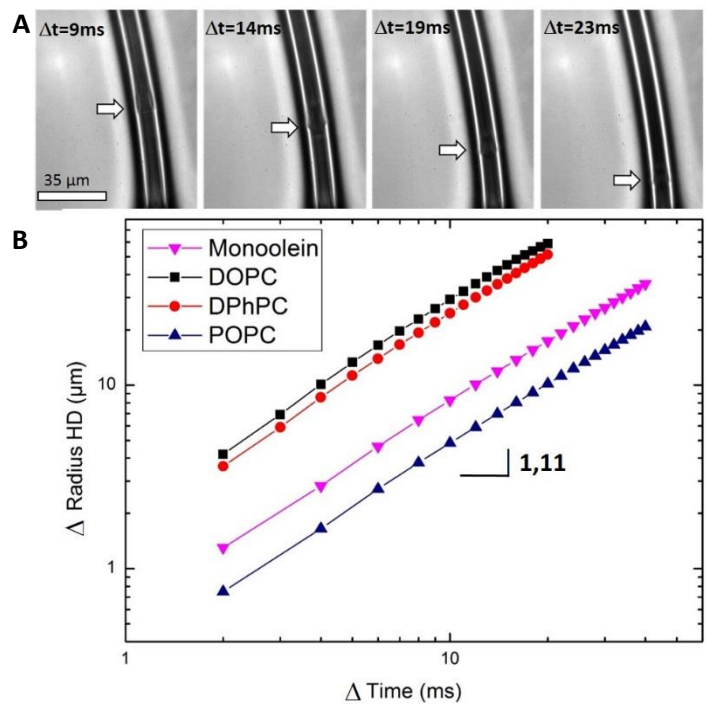

**Figure 6.** Fast formation of a stable hemifusion diaphragm. *Top (A):* Optical time series showing the evolution of two interacting DPhPC bilayers during a dewetting process. The center of the initially formed HD region is at the top of the shown micrographs and only the bottom half is shown. The white arrow indicates the moving contact line symmetrically surrounding the initially formed HD region. *Bottom (B):* Radius of the formed HD region as function of time. The contact line motion can be fitted by power laws with a power of 1.14 / 1.11 / 1.09 / 1.08 for DPhPC / DOPC / Monoolein and POPC, respectively.

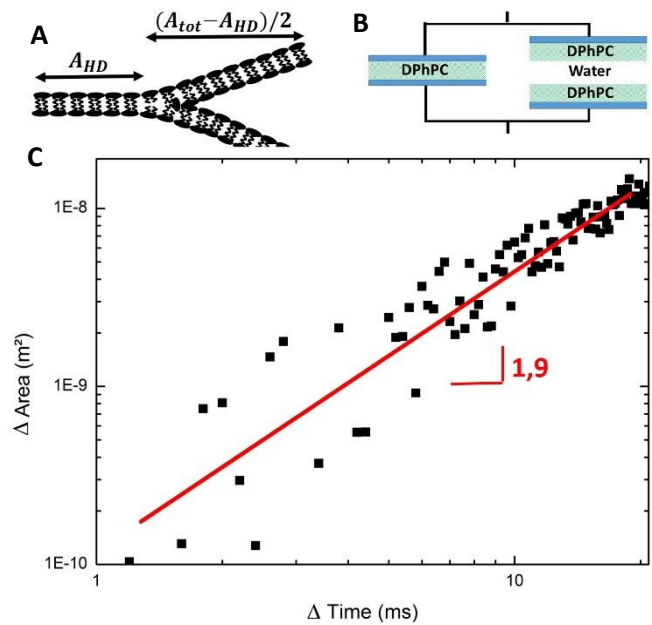

**Figure 7.** a) Schematic of the proposed dewetting mechanism leading to hemifusion. b) Schematic of the electrical equivalent circuit used to calculate the area increase from capacitance data shown in fig. 5. c) The increase of HD area $\Delta A_{HD}$ as a function of time $\Delta t$ calculated from the capacitance measurements during the expansion of a DPhPC hemifusion diaphragm. The evolution of $\Delta A_{HD}$ can be fitted with a power law of power 1.9.

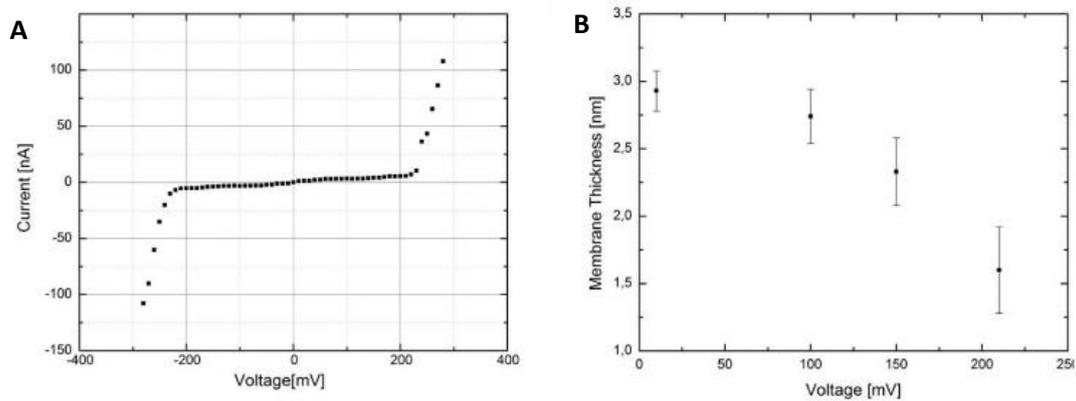

**Figure 8.** a) Current – voltage signal measured across a monoolein hemifusion diaphragm. b) Average thickness of the monoolein hemifused diaphragm as function of the applied voltage.

|  | **DOPC** | **DPhPC** | **POPC** | **Monoolein** |
|---|---|---|---|---|
| Surface tension $\gamma$ (water-oil interface) [mN/m] | 4,1 ±0,1 | 4 ±0,1 | 0,9 ±0,1 | 1,8 ±0,1 |
| Contact angle $\theta$ (single bilayer) [deg] | 22 ±1 | 23 ±1 | 23 ±1 | 23 ±1 |
| Membrane tension $\Gamma = 2\gamma \cos\theta$ (single bilayer) [mN/m] | 7,6 ±0,5 | 7,4 ±0,5 | 1,7 ±0,3 | 3,3 ±0,3 |
| zipping velocity d$L$/d$t$ (single bilayer) [µm/ms] | 0,9 ±0,1 | 6,1 ±0,1 | 0,3 ±0,1 | 0,5 ±0,1 |
| Contact angle $\theta_{HD}$ (hemifused state) [deg] | 55 ±1 | 54 ±1 | 54 ±1 | 56 ±1 |
| Membrane tension $\Gamma_{HD} = 2\Gamma \cos\theta_{HD}$ (hemifused state) [mN/m] | 8,7 ±0,8 | 8,5 ±0,8 | 1,9 ±0,3 | 3,7 ±0,4 |
| Velocity of HD growth d$R$/d$t$ (hemifused state) [µm/ms] | 2,9 ±0,2 | 2,6 ±0,2 | 0,5 ±0,2 | 0,9 ±0,2 |

**Table 1**. Summarized experimental values for surface tension $\gamma$, contact angle $\theta$, *zipping* velocity d$L$/d$t$, membrane tension $\Gamma$ for a single bilayer, membrane tension $\Gamma_{HD}$ for a hemifused state, and the velocity of the HD growth for a single state of hemifusion for DOPC, DPhPC, monoolein and POPC molecules.

## Conclusion

A microfluidic scheme is presented which is able to produce two unsupported lipid bilayers at desired location. The microfluidic scheme further allows to manipulate the two bilayers which can be brought into contact and thus to study the (dynamic) formation of hemifused states and to trigger single fusion events. The formation of bilayers and hemifusion diaphragms is shown for four different lipids whereas the existence of solvent free bilayer and hemifusion diaphragm are proofed by capacitance measurements. Simultaneous optical measurements revealed an ultra-fast formation process of a hemifusion diaphragm which is five orders of magnitude faster than previously observed for supported membranes and two orders of magnitude faster than previously observed for unilamellar vesicles. The formation process of a hemifusion diaphragm could be divided into an very fast initial symmetry breaking process where an initial hemifusion diaphragm of finite size (20 – 60) μm is formed followed by a growth of the hemifusion diaphragm which could be described by a dewetting process with no-slip boundary conditions. The corresponding hemifusion state formed by this process is equivalent to a dead-end or equilibrium state and no spontaneous fusion could be observed. Exploring the electrofusion of a monoolein hemifusion diaphragm, the emergence of reversible nanopores and a reduced stability compared to monoolein bilayer could be explored. The observed results escape from the results obtained for supported bilayer and emphasize the importance to study them in a free standing geometry. Moreover, the microfluidic scheme is quite flexible and allows to explore one or multi-component lipid membranes, with or without additional proteins or peptides and can be also fabricated in materials allowing e.g. for simultaneous x-ray analysis[29]. Thus we are convinced that the presented microfluidic scheme will stimulate further research into this area.

## Acknowledgements
The authors thank Marcus Müller, Karin Jacobs, Josh D. McGraw and Matthias Lessel for stimulating discussions.

## Notes and references

[a] Experimental Physics, Saarland University, 66123 Saarbrücken, Germany

[b] Max Planck Institute for Dynamics and Self-Organization, 37077 Goettingen, Germany

[*] E-mail: jean-baptiste.fleury@physik.uni-saarland.de

Electronic Supplementary Information (ESI) available:

Movie 1: Automated formation of two parallel lipid bilayers.

Movie 2: Automated manipulation of the two lipid bilayers to bring them into contact.

Movie 3: Visualization of a zipping process between two lipid monolayers to produce an oil-free lipid bilayer.

Movie 4a: Interaction of two lipid bilayers to produce a single state of hemifusion by a two-step process. This movie is presenting the first step of this process, i.e the emergence of a symmetry breaking.

Movie 4b: Interaction of two lipid bilayers to produce a single state of hemifusion by a two-step process. This movie is presenting the second step of this process, i.e the dewetting process.

Movie 5: Realization of a single fusion event after contacting two parallel lipid bilayers and formation of a state of hemifusion.